\documentclass[conference]{IEEEtran}
\IEEEoverridecommandlockouts
\usepackage{cite}
\usepackage{amsmath,amssymb,amsfonts}
\usepackage{algorithmic}
\usepackage{graphicx}
\usepackage{textcomp}
\usepackage{xcolor}
\usepackage{multirow}
\def\BibTeX{{\rm B\kern-.05em{\sc i\kern-.025em b}\kern-.08em
    T\kern-.1667em\lower.7ex\hbox{E}\kern-.125emX}}
\begin{document}

\title{Comparison Performance of Spectrogram and Scalogram as Input of Acoustic Recognition Task\\
}

\author{\IEEEauthorblockN{1\textsuperscript{st} Dang Thoai Phan}
\IEEEauthorblockA{\textit{Electrical Engineering Department} \\
\textit{BHT University of Applied Sciences and Technology, Berlin}\\
Berlin, Germany \\
thoai.phandang@gmail.com}
}

\maketitle

\begin{abstract}
Acoustic recognition has emerged as a prominent task in deep learning research, frequently utilizing spectral feature extraction techniques such as the spectrogram from the Short-Time Fourier Transform and the scalogram from the Wavelet Transform. However, there is a notable deficiency in studies that comprehensively discuss the advantages, drawbacks, and performance comparisons of these methods. This paper aims to evaluate the characteristics of these two transforms as input data for acoustic recognition using Convolutional Neural Networks. The performance of the trained models employing both transforms is documented for comparison. Through this analysis, the paper elucidates the advantages and limitations of each method, provides insights into their respective application scenarios, and identifies potential directions for further research.
\end{abstract}

\begin{IEEEkeywords}
spectrogram and scalogram, acoustic recognition, Convolutional Neural Networks
\end{IEEEkeywords}

\section{Introduction}
In the field of acoustic recognition, features in audio signals are extracted using spectral and time analysis techniques, most commonly the Short-Time Fourier Transform (STFT) and the Wavelet Transform (WT). The STFT decomposes signals in a linear frequency manner, while the WT decomposes time-domain signals into varying scales of frequency. The outputs of these transforms, known as spectrograms and scalograms, respectively, are subsequently input into deep learning models for classification tasks. Numerous studies have been conducted on this topic in recent years.

In the adoption of spectrograms as images for speech emotion recognition, Convolutional Neural Networks (CNNs) were employed to predict emotions in audio speech, as demonstrated in the study by \cite{badshah2017speech}. The model, comprising only three convolutional layers and three fully connected layers, outperformed the well-known pre-trained model AlexNet. A similar approach using CNNs and spectrograms was proposed in the study by \cite{santos2021audio} for acoustic event detection (AED) on audio recordings from a surveillance system. This method outperformed a benchmark AED-capable system. The study by \cite{purohit2019mimii} investigated abnormal machinery sound detection using spectrograms and an autoencoder. Two types of audio -- normal and abnormal sounds collected in a factory -- were transformed into spectrograms, which were then used for classification tasks to determine whether the machinery was in normal condition or damaged.

Scalograms, another audio feature extraction method, have seen increasing application in recent years. An implementation of CNNs to detect milling chatter using scalograms was developed in the study by \cite{tran2020milling}. Their system, designed for real-time detection, achieved superior performance compared to benchmark methods utilizing traditional machine learning. A similar approach to domestic audio classification was applied in the study by \cite{copiaco2019scalogram}. Scalograms were produced using the output of the continuous Wavelet Transform (CWT) and then fed into a pre-trained neural network to predict whether the sound originated from social activities or a vacuum cleaner. Utilizing scalograms and CNNs for audio scene modeling, the research by \cite{chen2018deep} demonstrated better performance for the model trained with scalograms compared to one using spectrograms.

As illustrated above, while numerous studies have explored audio recognition using spectrograms or scalograms in conjunction with deep learning, few have directly compared the performance of these two feature extraction methods. Existing benchmarks often offer indirect comparisons, as they involve different prediction models rather than a comprehensive evaluation of the feature extractors themselves, as seen in the study by \cite{chen2018deep}. Or the study by \cite{huzaifah2017comparison} compared different transforms but focused solely on predictive performance, lacking a comparative analysis of how the attributes of the transforms influence predictive performance and computational expense. The research by \cite{tzanetakis2001audio} compared various time-frequency transforms but used discrete coefficients directly instead of scalograms or spectrograms, limiting the models’ performance. Given these limitations, this paper aims to develop an approach for a rigorous comparison of spectrograms and scalograms as audio feature extractors. To this end, the experimental design for both methods is kept identical, including the original audio data used, the format of input data, the configuration of the CNNs models, and the training and evaluation methods. The sole distinction in this study is the type of feature extractor: spectrogram versus scalogram.
\section{Theoretical foundation}

\subsection{Short-time Fourier transform}

The STFT \cite{muller2015fundamentals} is a method used to analyze signals by dividing them into segments based on time frames and performing Fourier transforms (FT) on each segment, denoted as (\ref{STFT equation}). This approach allows the signal to be examined in both the time shift \begin{math}\tau\end{math} and frequency shift \begin{math}\omega\end{math} simultaneously. By applying a windowing function, denoted as \begin{math}\gamma(t)\end{math}, to isolate specific portions of the signal, STFT computes the spectrum over short durations, enabling it to capture not only the frequencies present in the signal but also how these frequencies change over time.
\begin{equation}
X _{STFT}(\tau, \omega) = \int_{-\infty}^{\infty} \,x(t)\gamma^*(t-\tau)e^{-j\omega t}\,dt   
\label{STFT equation}
\end{equation}
\begin{equation}
X _{STFT}(m, k) = \sum_{n=0}^{N-1} \,x(n)\gamma^*(n-mH)e^{-j2\pi kn/N}
\label{discrete STFT equation}
\end{equation}

In practical applications, signals are processed in discrete time formats, and the STFT is computed using a window of length N, referred to as the frame size, denoted as (\ref{discrete STFT equation}). Since shifting sample-wise results in redundant data due to the similarity of neighboring spectra, a larger step size, known as the hop size H, is used, typically set to half the window length H=N/2. This choice strikes a balance between achieving high time resolution and managing data volume effectively \cite{muller2015fundamentals}. Each sample point in the data sequence, where STFT is calculated, corresponds to a frame index m defined as m=M/H, with M representing the total data length. Within each frame, computing the FT of a discrete signal of N points in time results in a frequency vector of N points. Due to the symmetric nature of the Fourier spectrum, only the first half of the spectrum (k=0 to N/2) is typically considered to avoid redundancy. Consequently, the output of STFT is structured as a matrix with dimensions (M/H, N/2+1) where each element X(m,k) represents the complex Fourier coefficient of the \(m^{th}\) time frame at frequency index k.

\begin{figure}[t]
\centering
\includegraphics[width=0.45\textwidth]{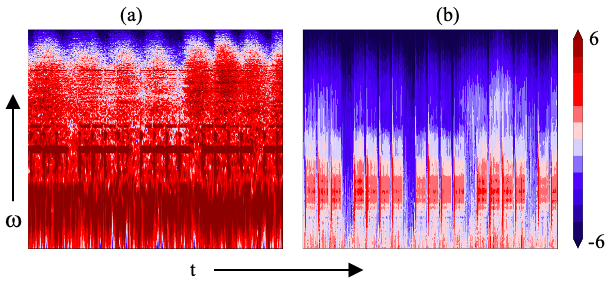}
\caption{Visualization of spectrogram (a) and scalogram (b)}
\label{VisualizationSpecScaloFig}
\end{figure}

To enhance the visualization of STFT results, spectrograms are commonly employed \cite{mertins1999signal}. Represented as a type of heatmap image, they provide a two-dimensional representation of signal energy across time and frequency dimensions. Time is arranged along the x-axis, and frequency along the y-axis. The intensity of color in each pixel of the spectrogram corresponds to the degree of signal energy. For instance, the colors range from dark blue, indicating lower energy, to dark red, indicating higher energy, as illustrated in Fig.~\ref{VisualizationSpecScaloFig}(a).

\subsection{Wavelet transform}
WT is a technique used to transform a signal into a representation that enhances specific features for further processing \cite{addison2017illustrated}. It decomposes a signal from the time domain into the time-frequency domain using a mother wavelet, denoted as (\ref{CWT equation}). The WT depends on parameters of time shift \textit{b} and frequency scale \textit{a}. The normalization factor 1/\begin{math}\sqrt{\textit{a}}\end{math} ensures energy normalization across scales. During transformation, the mother wavelet \begin{math}\psi(t)\end{math} undergoes dilation and contraction as the scale \textit{a} varies, and then slides along the time axis with step \textit{b} to convolve with the signal x(t), generating a matrix of coefficients.
\begin{equation}
X _{WT}(\textit{b}, \textit{a}) = \frac{1}{\sqrt{\textit{a}}}\int_{-\infty}^{\infty} \,x(t)\psi^*(\frac{t-\textit{b}}{\textit{a}})\,dt
\label{CWT equation}
\end{equation}
\begin{equation}
X _{WT}(m, k) = \frac{1}{\sqrt{\textit{a}}}\sum_{n=0}^{N-1} \,x(n)\psi^*(\frac{n-k}{m})
\label{discrete CWT equation}
\end{equation}
The WT for discretized signals is computed by discretizing the transform, replacing the integral with a discrete summation of values within the sampling interval, denoted as (\ref{discrete CWT equation}). Both the translation parameter and the scale parameter are also in discrete forms. For the CWT, the translation occurs sample-wise, producing a matrix with dimensions (N, \textit{a}), where N represents the data length and \textit{a} represents the scale range. In contrast, the Discrete WT (DWT) is implemented using filter banks, which are computationally less expensive than the CWT. However, the result of the DWT is not a matrix, making it unsuitable for generating heat maps.

Scalogram is calculated as squared magnitude of the WT. Similar to a spectrogram, a scalogram is also used to visualize the distribution of signal energy across time and frequency dimensions, represented as a heat map with intensity expressed through a range of colors, as illustrated in Fig.~\ref{VisualizationSpecScaloFig}(b).

\subsection{Time and frequency resolution of transforms}\label{AA}
\begin{figure}[t]
\centering
\includegraphics[width=0.45\textwidth]{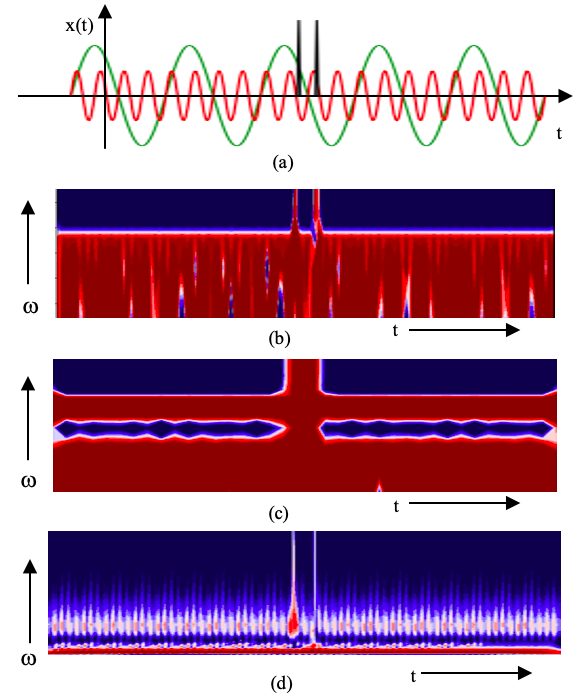}
\caption{Time and frequency resolution of spectrogram and scalogram}
\label{MultiresolutionFig}
\end{figure}

\paragraph{Uncertainty principle} To achieve a comprehensive analysis of a signal, both time resolution and frequency resolution should be maximized. This implies capturing signal variations with minimal shifts in both time and frequency. However, the uncertainty principle \cite{mertins1999signal} imposes a constraint on the tradeoff between time and frequency resolution, as described in (\ref{uncertainty equation}). Utilizing a narrow time window (small \begin{math} \Delta _t \end{math}) allows for high time resolution in STFT or WT, but results in poor frequency resolution (large \begin{math} \Delta _\omega \end{math}). Conversely, employing a large time window yields poor time resolution but improved frequency resolution.
\begin{equation}
\Delta _t.\Delta _\omega \geq \frac{1}{2}
\label{uncertainty equation}
\end{equation}
\paragraph{Multiresolution} Fig.~\ref{MultiresolutionFig} provides a visual comparison of the WT via scalogram and the STFT via spectrogram. As depicted, the time signal x(t) contains two periodic components and two Dirac impulses, as shown in Fig.~\ref{MultiresolutionFig}(a). The spectrogram of x(t) with a short window, illustrated in Fig.~\ref{MultiresolutionFig}(b), offers high time resolution but low frequency resolution. Consequently, the two distinctive impulses are clearly visible at the center, whereas the two components in the frequency axis are merged. Fig.~\ref{MultiresolutionFig}(c) shows the spectrogram with a long window, which provides low time resolution but high frequency resolution. This results in two clearly separated red horizontal stripes indicating the frequency components, but the two impulses are merged at the center. In contrast, the scalogram in Fig.~\ref{MultiresolutionFig}(d) effectively represents both the two distinctive impulses and the two separate red horizontal stripes. This capability of the scalogram is attributed to the multiresolution feature of the WT, which achieves high time resolution at high frequencies and high frequency resolution at low frequencies.
\section{Experiment}
\subsection{Workflow}
\begin{figure}[t]
\centering
\includegraphics[width=0.45\textwidth]{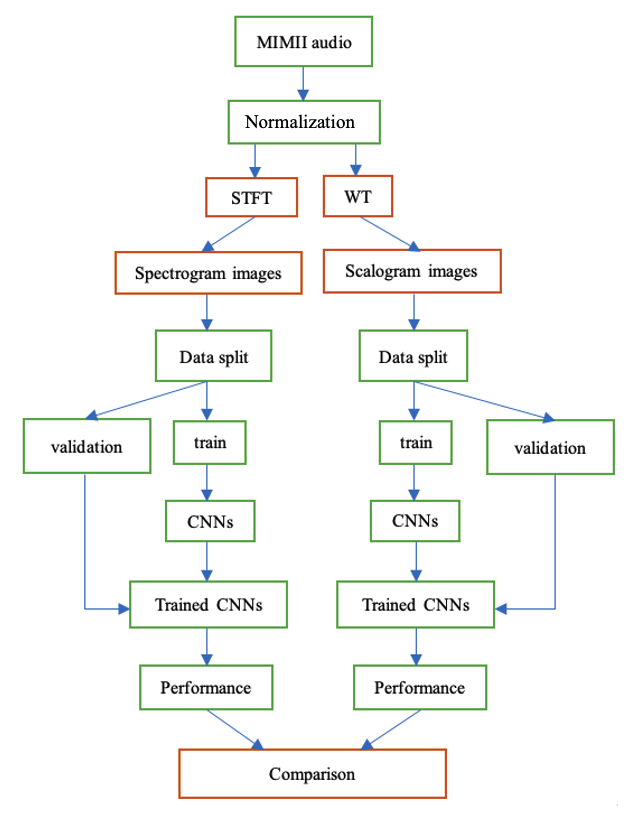}
\caption{Experimental Workflow}
\label{WorkflowFig}
\end{figure}
The entire workflow of the experiment in this study is presented in Fig.~\ref{WorkflowFig}. Initially, the audio data is normalized to constrain the amplitude values within the range (0, 1) before being transformed using the STFT and WT, respectively. Following the transformations, the matrices produced as outputs from the STFT and WT are used to create spectrograms and scalograms. These are then plotted as images and split into training and validation sets for a CNNs model. The performance of the trained CNNs is then recorded and compared between the two types of data, using metric the Area Under the Curve of the Receiver Operating Characteristic (AUC-ROC) \cite{provost1998case}.
\subsection{MIMII audio dataset}
The acoustic recognition task in this study is conducted using the MIMII audio dataset for machinery fault detection \cite{purohit2019mimii}. This dataset was created to provide real-life sound data from factories for various machine types, including fans, pumps, sliders, and valves. These machines produce both stationary and non-stationary sounds, presenting varying levels of difficulty in distinguishing between normal and abnormal conditions. Background noise from various factory environments is recorded and mixed with the target machine sounds to simulate real-world conditions at three levels of signal-to-noise ratio (SNR): -6 dB, 0 dB, and 6 dB. Each audio recording has a length of 10 seconds, sampled at 16 kHz, resulting in each discrete audio 160,000 samples.
\subsection{Audio normalization}\label{SCM}
\begin{figure}[t]
\centering
\includegraphics[width=0.45\textwidth]{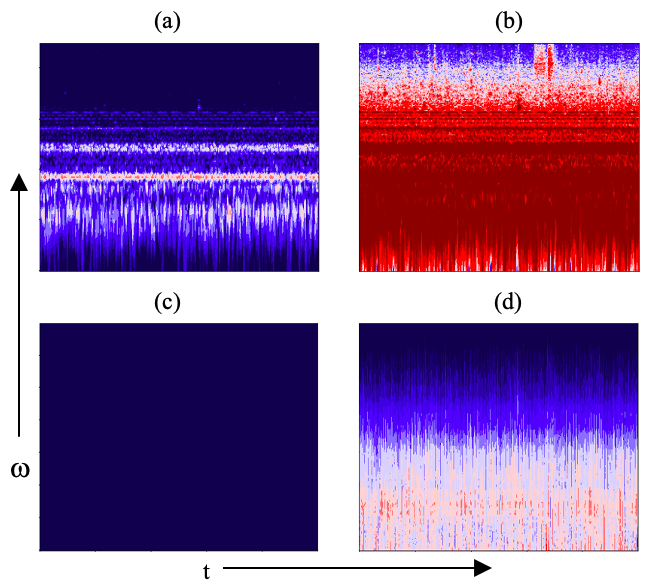}
\caption{Effect of normalization technique}
\label{NormalizationFig}
\end{figure}
\begin{equation}
y _n=\frac{y _n}{max|y _n|}
\label{Normalization equation}
\end{equation}
In factories, the recording conditions of audio data often vary in reality. Consequently, the amplitude of recorded audio can differ for the same type of error across different situations, or even within the same situation if the microphone position or vibration changes. For instance, placing the microphone closer results in higher recorded amplitudes, while placing it further away results in lower amplitudes. To mitigate the effects of these variations, this paper employs normalization step to constrain amplitudes of signals within the range (0, 1). The maximum amplitude of a signal, denoted as \begin{math}max(|y _{n}|)\end{math}, is computed, and all other sample values are normalized by this maximum value, as described in (\ref{Normalization equation}). This step ensures that if the maximum amplitude of a signal exceeds 1, it is scaled down to 1; conversely, if it is below 1, it is scaled up to 1. It aims to make the structures of audio signals of the same error type more uniform and to accentuate differences between signals of different error types. Indeed, the normalization significantly enhances the visibility of the generated spectrogram and scalogram, as illustrated in Fig. \ref{NormalizationFig}. Where no normalization is applied as in Fig.\ref{NormalizationFig} (a) and (c), it is challenging for human observers to discern the heat maps of the spectrogram and scalogram. In contrast, in Fig.\ref{NormalizationFig} (b) and (d), where normalization is applied, the heat maps are visualized much more clearly.
\subsection{Implementation of Short-Time Fourier Transform}
The STFT is conducted in a time-discrete manner using Librosa \cite{mcfee2015librosa}. Initially, the frame size N and hop size H are treated as hyperparameters aimed at optimizing the performance of the trained model during the initial stages of training. Subsequently, a configuration of STFT with N = 1024 and H = 512 demonstrating good predictive performance is selected, resulting a matrix with dimensions (513, 313). This configuration is then applied uniformly across the entire audio dataset to generate the STFT representations.

\subsection{Implementation of Continuous Wavelet Transform}
The CWT is also executed in a time-discrete manner using Pywavelets \cite{lee2019pywavelets}. The translation is performed sample-wise, and the scale parameter encompasses a range of consecutive natural numbers. Given that the scale determines the frequency resolution, various scales are evaluated during the training of the CNNs model. Then the scale from 2 to 129, demonstrating optimal predictive performance, is selected, producing output matrices with dimensions (16000, 128). Subsequently, this selected scale is employed to generate scalograms for the entire audio dataset.

\subsection{Implementation of Convolutional Neural Networks}
The primary focus of this study is to compare the performance between spectrograms and scalograms. Therefore, the CNNs model is designed in an exploratory manner, ensuring it achieves satisfactory performance. The Tensorflow \cite{tensorflow2015-whitepaper} is used for CNNs implementation. Several configurations of CNNs are evaluated, and an effective configuration is chosen based on its performance, as outlined in Table \ref{CONFIGURATIONCNNsTab}.
\begin{table}[htbp]
\centering
\caption{Configuration of CNNs}
\label{CONFIGURATIONCNNsTab}
\begin{tabular}{|c|c|}
\hline
\textbf{Layer} & \textbf{Name}                                                                                                         \\ \hline
1              & Rescaling (1/255)                                                                                                     \\ \hline
2              & \begin{tabular}[c]{@{}c@{}}Conv2D(16, 3, activation='relu', use\_bias=True,\\ bias\_initializer='zeros')\end{tabular} \\ \hline
3              & MaxPooling2D                                                                                                          \\ \hline
4              & \begin{tabular}[c]{@{}c@{}}Conv2D(32, 3, activation='relu', use\_bias=True,\\ bias\_initializer='zeros')\end{tabular} \\ \hline
5              & MaxPooling2D                                                                                                          \\ \hline
6              & \begin{tabular}[c]{@{}c@{}}Conv2D(64, 3, activation='relu', use\_bias=True,\\ bias\_initializer='zeros')\end{tabular} \\ \hline
7              & MaxPooling2D                                                                                                          \\ \hline
8              & Flatten                                                                                                               \\ \hline
9              & Dense(128, activation='relu')                                                                                         \\ \hline
10             & Dropout(0.25)                                                                                                         \\ \hline
11             & Dense(256, activation='relu')                                                                                         \\ \hline
12             & Dropout(0.25)                                                                                                         \\ \hline
13             & Dense(units=2, activation='softmax')                                                                                  \\ \hline
\end{tabular}
\end{table}

\section{Benchmarking and performance evaluation}
\subsection{Benchmark study}
The benchmark study \cite{gantert2021supervised}, representing the state-of-the-art for supervised methods on the same dataset, employed Mel-frequency cepstral coefficients (MFCCs) as spectral features. Unlike the approach of transforming data into compressed images, this model directly utilized tuples generated by MFCCs. Prior to being used for training, the tuples were normalized to have zero mean and unit standard deviation. The prediction model used in this study was a Multilayer Perceptron (MLP). Additionally, the tuples of abnormal sounds were oversampled to achieve a balanced dataset. Detailed comparisons are presented in Table \ref{BENCHMARKTab}.
\begin{table}[]
\centering
\caption{Comparison with benchmark study}
\label{BENCHMARKTab}
\begin{tabular}{|c|ccc|}
\hline
\multicolumn{1}{|l|}{} & \multicolumn{1}{c|}{\textbf{Baseline}} & \multicolumn{1}{c|}{\textbf{Spectrogram}} & \textbf{Scalogram} \\ \hline
\textbf{Method}           & \multicolumn{3}{c|}{Supervised learning}                                                \\ \hline
\textbf{Metric}           & \multicolumn{3}{c|}{AUC\_ROC}                                                           \\ \hline
\textbf{Normalization}    & \multicolumn{1}{c|}{tuples $\sim$N(0,1)} & \multicolumn{2}{l|}{audio $y_n=y_n/max(|y_n|)$}                \\ \hline
\textbf{Spectral feature} & \multicolumn{1}{c|}{MFCCs}               & \multicolumn{1}{c|}{Spectrogram} & Scalogram \\ \hline
\textbf{Input of model}   & \multicolumn{1}{c|}{tuples}              & \multicolumn{2}{c|}{images}                  \\ \hline
\textbf{Oversampling}     & \multicolumn{1}{c|}{yes}                 & \multicolumn{2}{c|}{no}                      \\ \hline
\textbf{Model}            & \multicolumn{1}{c|}{MLP}                 & \multicolumn{2}{c|}{CNNs}                    \\ \hline
\end{tabular}
\end{table}

\begin{table}[htbp]
\centering
\caption{Comparison of computational expense}
\label{COMPUTATIONAL EXPENSE Tab}
\begin{tabular}{|l|ccc|}
\hline
\multirow{2}{*}{} & \multicolumn{3}{c|}{\textbf{Single file}}   \\ \cline{2-4} 
                                    & \multicolumn{1}{c|}{\textbf{Spectrogram}} & \multicolumn{1}{c|}{\textbf{Scalogram}} & \textbf{Deviation} \\ \hline
\multicolumn{1}{|c|}{\textbf{Time}} & \multicolumn{1}{c|}{0.58}                 & \multicolumn{1}{c|}{22.38}              & 21.8               \\ \hline
                  & \multicolumn{3}{c|}{\textbf{Whole dataset}} \\ \hline
\multicolumn{1}{|c|}{\textbf{Time}} & \multicolumn{1}{c|}{10,451.02}            & \multicolumn{1}{c|}{392,814.2}          & 392,814.2          \\ \hline
\end{tabular}
\end{table}

\begin{table}[htbp]
\centering
\caption{Performance benchmarking}
\label{PERFORMANCE BENCHMARKING Tab}
\begin{tabular}{|c|c|c|c|}
\hline
               & \textbf{baseline} & \textbf{scalogram} & \textbf{spectrogram} \\ \hline
\textbf{-6 dB} & 0.893               & 0.921              & 0.981                  \\ \hline
\textbf{0 dB}  & 0.942             & 0.964                & 0.992                  \\ \hline
\textbf{6 dB}  & 0.975             & 0.988              & 0.997                  \\ \hline
\end{tabular}
\end{table}

\begin{table}[htbp]
\centering
\caption{Spectrogram and scalogram comparison}
\label{PERFORMANCE COMPARISON BETWEEN TWO TRANSFORMS Tab}
\begin{tabular}{|c|c|c|}
\hline
                & \textbf{scalogram} & \textbf{spectrogram} \\ \hline
\textbf{fan}    & 0.934              & 0.988                \\ \hline
\textbf{pump}   & 0.962              & 0.991                \\ \hline
\textbf{slider} & 0.947              & 0.995                \\ \hline
\textbf{valve}  & 0.987              & 0.984                \\ \hline
\end{tabular}
\end{table}

\begin{figure}[t]
\centering
\includegraphics[width=0.45\textwidth]{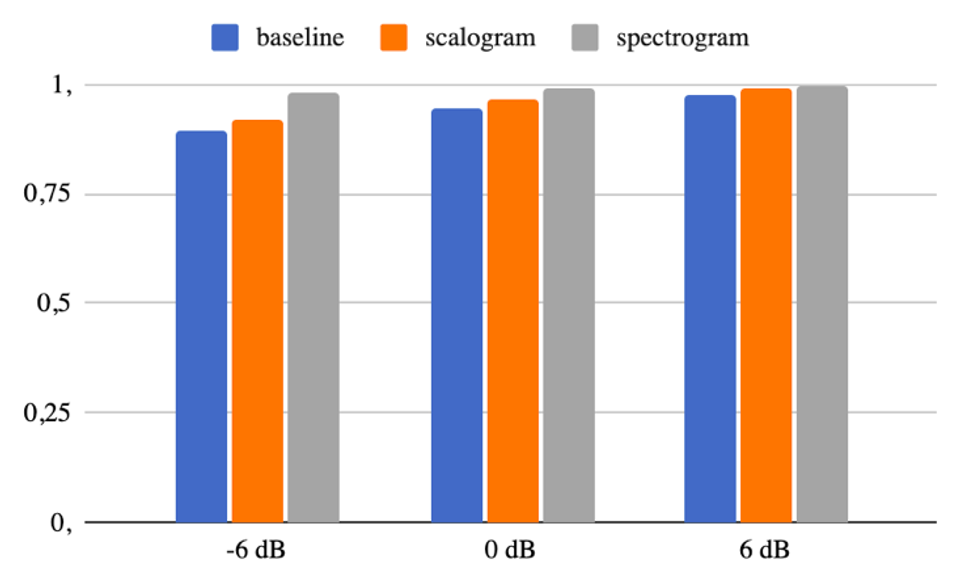}
\caption{Performance benchmarking}
\label{BenchmarkingFigure}
\end{figure}

\begin{figure}[t]
\centering
\includegraphics[width=0.45\textwidth]{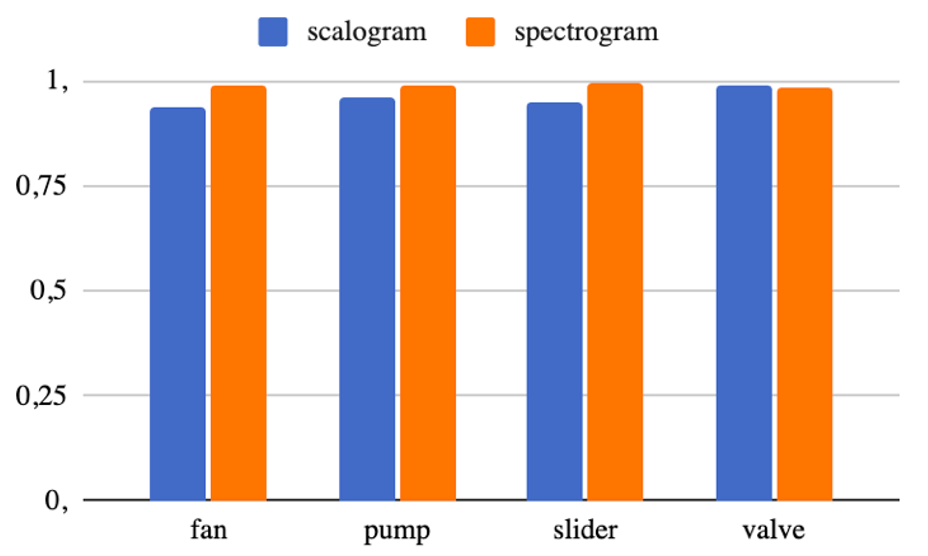}
\caption{Performance comparison between two transforms}
\label{ComparisionPerformanceFig}
\end{figure}
\subsection{Performance evaluation}
\paragraph{Computational expense} Due to differences in the computation methods of the STFT and the CWT, the resulting output matrices vary in size, necessitating a comparison of computational efficiency. For simplicity, computational expense is evaluated by measuring the time (in seconds) required for spectrogram and scalogram generation using the hardware employed in this study. As shown in Table \ref{COMPUTATIONAL EXPENSE Tab}, scalogram generation exhibits significantly higher computational expense compared to spectrogram generation. The time deviation for generating a single file is 21.8 seconds, and for processing the entire dataset of 18,019 audio files, it totals 392,814.2 seconds. Generating scalograms for the entire dataset requires approximately 109 hours, whereas generating spectrograms takes approximately 2.9 hours. This substantial difference arises because the CWT computes coefficients for every single sample among the 160,000 samples, whereas the STFT computes them only at intervals defined by the hop size.
\paragraph{Predictive performance benchmarking} The performance of CNNs models applied to spectrograms and scalograms is assessed by averaging results over ten training and evaluation runs. The benchmark results are summarized in Table \ref{PERFORMANCE BENCHMARKING Tab} and visually represented in Figure \ref{BenchmarkingFigure}. As depicted, the averaged predictive performance of the models consistently improves with increasing SNR for both spectrograms and scalograms, as well as the baseline. This correlation between SNR and performance is evident because higher SNR enhances the clarity of normal and abnormal machine sounds, thereby facilitating more accurate classification. The spectrogram and scalogram designs in this study consistently outperform those of previous research across three SNR levels. This indicates that the spectrogram and scalogram designs have achieved relatively high performance standards.
\paragraph{Spectrogram and scalogram predictive performance comparision}
For the comparison between spectrogram and scalogram within each machine type, the results are presented in Table \ref{PERFORMANCE COMPARISON BETWEEN TWO TRANSFORMS Tab} and visualized in Figure \ref{ComparisionPerformanceFig}. Overall, spectrograms consistently outperform scalograms, with the exception of one case involving valves where CNNs predicts audio based on scalograms more effectively than those based on spectrograms. This exception arises due to the non-stationary nature of valve audio signals \cite{purohit2019mimii}, which are impulsive and sparse in time. The WT used in scalogram generation leverages its multiresolution feature to enhance feature extraction for such non-stationary signals \cite{mertins1999signal}. Conversely, when evaluating performance on fan audio signals, spectrograms achieve significantly higher performance compared to scalograms. This difference can be attributed to the stationary nature of fan sounds \cite{purohit2019mimii}, where the constant time and frequency resolution of spectrograms are more adept at feature extraction than scalograms.
\section{Conclusion and future work}
\subsection{Conclusion}
This paper implements a comparative analysis of two acoustic recognition approaches, focusing on the spectrogram and scalogram as audio feature extractors, and evaluating their respective designs. The classification task for distinguishing between normal and abnormal sounds is successfully conducted using CNNs. By employing arbitrary CNNs designs, the configurations of spectrogram and scalogram consistently demonstrate superior performance compared to benchmark methods. The study fulfills its primary objective by providing a comprehensive comparison between spectrogram and scalogram in terms of their characteristics, transform configurations, and model performance evaluations. Overall, the STFT design in this paper generally outperforms the WT, with the exception of non-stationary valve audio. For stationary fan audio, STFT significantly outperforms WT in predictive performance.
\subsection{Future work}
When configuring the STFT and CWT, it is observed that their output matrices differ in size due to the computational requirements. CWT necessitates computation for every single sample of data, whereas STFT computes values at intervals defined by the hop size. Consequently, this disparity may limit the direct comparability between the two transforms. It suggests a potential avenue for future research, where the translation parameter and scale of the CWT are adjusted to produce output matrices of sizes comparable to STFT. Achieving this alignment could lead to more balanced computational expenses and a rigorous comparison between STFT and CWT. Moreover, the current PyWavelets library facilitates scale tuning for CWT but lacks support for adjusting the translation parameter. Therefore, developing a library that enables tuning of both parameters in CWT to reduce computational costs is a promising direction for future advancements.

The experiment in this paper employs a normalization technique that confines signal amplitudes to within the range of 0 to 1, which proved advantageous for preprocessing audio signals before conducting STFT and WT analyses. However, alternative normalization methods, such as normalizing signal to zero mean and unit variance, have not been evaluated to determine their comparative effectiveness. Therefore, conducting experiments to compare different normalization techniques and assess their robustness, or applying each technique to different datasets to evaluate their general applicability, would represent a promising area of research. Such investigations could provide insights into optimizing signal preprocessing methodologies for various analytical tasks involving audio data.

In the discussion section of the performance comparison, scalograms demonstrate superior performance over spectrograms when the audio signal exhibits non-stationarity, attributed to their multiresolution capability. Conversely, spectrograms exhibit better predictive performance for stationary audio signals, leveraging their linear resolution. Thus, future research exploring how the stationary characteristics of signals influence the predictive performance of models using spectrograms and scalograms would be valuable. A potential research direction could involve deriving a function that quantifies the impact of signal stationarity on CNNs model performance. This task would parallel the role of SNR in influencing CNNs performance observed in this study, thereby enhancing understanding of spectrogram and scalogram effectiveness across varying signal conditions.

\bibliographystyle{unsrt}
\bibliography{paper1}

@inproceedings{badshah2017speech,
  title={Speech emotion recognition from spectrograms with deep convolutional neural network},
  author={Badshah, Abdul Malik and Ahmad, Jamil and Rahim, Nasir and Baik, Sung Wook},
  booktitle={2017 international conference on platform technology and service (PlatCon)},
  pages={1--5},
  year={2017},
  organization={IEEE}
}

@article{santos2021audio,
  title={Audio Attacks and Defenses against AED Systems--A Practical Study},
  author={Santos, Rodrigo dos and Nilizadeh, Shirin},
  journal={arXiv preprint arXiv:2106.07428},
  year={2021}
}

@article{purohit2019mimii,
  title={MIMII Dataset: Sound dataset for malfunctioning industrial machine investigation and inspection},
  author={Purohit, Harsh and Tanabe, Ryo and Ichige, Kenji and Endo, Takashi and Nikaido, Yuki and Suefusa, Kaori and Kawaguchi, Yohei},
  journal={arXiv preprint arXiv:1909.09347},
  year={2019}
}

@article{tran2020milling,
  title={Milling chatter detection using scalogram and deep convolutional neural network},
  author={Tran, Minh-Quang and Liu, Meng-Kun and Tran, Quoc-Viet},
  journal={The International Journal of Advanced Manufacturing Technology},
  volume={107},
  number={3},
  pages={1505--1516},
  year={2020},
  publisher={Springer}
}

@inproceedings{copiaco2019scalogram,
  title={Scalogram neural network activations with machine learning for domestic multi-channel audio classification},
  author={Copiaco, Abigail and Ritz, Christian and Fasciani, Stefano and Abdulaziz, Nidhal},
  booktitle={2019 IEEE International Symposium on Signal Processing and Information Technology (ISSPIT)},
  pages={1--6},
  year={2019},
  organization={IEEE}
}

@inproceedings{chen2018deep,
  title={Deep Convolutional Neural Network with Scalogram for Audio Scene Modeling.},
  author={Chen, Hangting and Zhang, Pengyuan and Bai, Haichuan and Yuan, Qingsheng and Bao, Xiuguo and Yan, Yonghong},
  booktitle={Interspeech},
  pages={3304--3308},
  year={2018}
}

@book{muller2015fundamentals,
  title={Fundamentals of music processing: Audio, analysis, algorithms, applications},
  author={M{\"u}ller, Meinard},
  volume={5},
  year={2015},
  publisher={Springer}
}

@book{mertins1999signal,
  title={Signal analysis: wavelets, filter banks, time-frequency transforms and applications},
  author={Mertins, Alfred and Mertins, Dr Alfred},
  year={1999},
  publisher={John Wiley \& Sons, Inc.}
}

@book{addison2017illustrated,
  title={The illustrated wavelet transform handbook: introductory theory and applications in science, engineering, medicine and finance},
  author={Addison, Paul S},
  year={2017},
  publisher={CRC press}
}

@inproceedings{mcfee2015librosa,
  title={librosa: Audio and music signal analysis in python.},
  author={McFee, Brian and Raffel, Colin and Liang, Dawen and Ellis, Daniel PW and McVicar, Matt and Battenberg, Eric and Nieto, Oriol},
  booktitle={SciPy},
  pages={18--24},
  year={2015}
}

@article{lee2019pywavelets,
  title={PyWavelets: A Python package for wavelet analysis},
  author={Lee, Gregory and Gommers, Ralf and Waselewski, Filip and Wohlfahrt, Kai and O'Leary, Aaron},
  journal={Journal of Open Source Software},
  volume={4},
  number={36},
  pages={1237},
  year={2019},
  publisher={The Open Journal}
}

@misc{tensorflow2015-whitepaper,
title={ {TensorFlow}: Large-Scale Machine Learning on Heterogeneous Systems},
url={https://www.tensorflow.org/},
note={Software available from tensorflow.org},
author={
    Mart\'{i}n~Abadi and
    Ashish~Agarwal and
    Paul~Barham and
    Eugene~Brevdo and
    Zhifeng~Chen and
    Craig~Citro and
    Greg~S.~Corrado and
    Andy~Davis and
    Jeffrey~Dean and
    Matthieu~Devin and
    Sanjay~Ghemawat and
    Ian~Goodfellow and
    Andrew~Harp and
    Geoffrey~Irving and
    Michael~Isard and
    Yangqing Jia and
    Rafal~Jozefowicz and
    Lukasz~Kaiser and
    Manjunath~Kudlur and
    Josh~Levenberg and
    Dandelion~Man\'{e} and
    Rajat~Monga and
    Sherry~Moore and
    Derek~Murray and
    Chris~Olah and
    Mike~Schuster and
    Jonathon~Shlens and
    Benoit~Steiner and
    Ilya~Sutskever and
    Kunal~Talwar and
    Paul~Tucker and
    Vincent~Vanhoucke and
    Vijay~Vasudevan and
    Fernanda~Vi\'{e}gas and
    Oriol~Vinyals and
    Pete~Warden and
    Martin~Wattenberg and
    Martin~Wicke and
    Yuan~Yu and
    Xiaoqiang~Zheng},
  year={2015},
}

@inproceedings{provost1998case,
  title={The case against accuracy estimation for comparing induction algorithms.},
  author={Provost, Foster J and Fawcett, Tom and Kohavi, Ron and others},
  booktitle={ICML},
  volume={98},
  pages={445--453},
  year={1998}
}

@inproceedings{gantert2021supervised,
  title={A supervised approach for corrective maintenance using spectral features from industrial sounds},
  author={Gantert, Luana and Sammarco, Matteo and Detyniecki, Marcin and Campista, Miguel Elias M},
  booktitle={2021 IEEE 7th World Forum on Internet of Things (WF-IoT)},
  pages={723--728},
  year={2021},
  organization={IEEE}
}

@article{huzaifah2017comparison,
  title={Comparison of time-frequency representations for environmental sound classification using convolutional neural networks},
  author={Huzaifah, Muhammad},
  journal={arXiv preprint arXiv:1706.07156},
  year={2017}
}

@inproceedings{tzanetakis2001audio,
  title={Audio analysis using the discrete wavelet transform},
  author={Tzanetakis, George and Essl, Georg and Cook, Perry},
  booktitle={Proc. conf. in acoustics and music theory applications},
  volume={66},
  year={2001},
  organization={Citeseer}
}
\vspace{12pt}
\end{document}